\documentclass[%
aip,jcp,amsmath,amssymb, reprint, twocolumn
]{revtex4-2}
\usepackage{amsmath}
\usepackage{mathtools}
\usepackage{float}
\usepackage{amsfonts}
\usepackage{amssymb}
\usepackage{dcolumn} %% tables cols aligned at decimal point
\usepackage{array}
\newcolumntype{P}[1]{>{\centering\arraybackslash}p{#1}}
\newcolumntype{M}[1]{>{\centering\arraybackslash}m{#1}}
\newcolumntype{C}[1]{>{\centering\arraybackslwash}p{#1}}
\usepackage{float}
\usepackage{psfrag}
\usepackage{stackengine}
\usepackage{amssymb}
\usepackage{mathtools}  
\usepackage{xfrac} 
\usepackage[T1]{fontenc}
\usepackage{graphicx}
\usepackage{diagbox}
\usepackage{tikz}
\usetikzlibrary{positioning}
\usetikzlibrary{arrows}
\usetikzlibrary{trees}
\usepackage{amssymb}
\usetikzlibrary{decorations.pathmorphing}
\usetikzlibrary{decorations.markings}
\usetikzlibrary{automata,positioning}
\usepackage{braket}
\usepackage{hyperref}
\usepackage{rotating}
\usepackage{adjustbox}
\usepackage{ragged2e}
\usepackage{simplewick}
\usepackage{simpler-wick}
\usepackage{csquotes}
\usepackage{physics,amsmath}
\usepackage{xcolor}
\usepackage{fancyhdr}
\usepackage{csquotes}
\UseRawInputEncoding
\usepackage{soul}
\begin{document}

\author{Chayan Patra}
\altaffiliation{Contributed equally to the work}
\affiliation{ Department of Chemistry,\\ Indian Institute of Technology Bombay,\\ Powai, Mumbai, PIN: 400076, India}
\author{Valay Agarawal}
\altaffiliation{Contributed equally to the work}
\affiliation{ Department of Chemistry,\\ University of Chicago, \\ Chicago, IL 60637, USA}
\author{Dipanjali Halder}
\affiliation{ Department of Chemistry,\\ Indian Institute of Technology Bombay,\\ Powai, Mumbai, PIN: 400076, India}
\author{Anish Chakraborty}
\affiliation{ Department of Chemistry,\\ Indian Institute of Technology Bombay,\\ Powai, Mumbai, PIN: 400076, India}
\author{Dibyendu Mondal}
\thanks{Authors contributed equally to the work}
\affiliation{ Department of Chemistry,\\ Indian Institute of Technology Bombay,\\ Powai, Mumbai, PIN: 400076, India}
\author{Sonaldeep Halder}
\thanks{Authors contributed equally to the work}
\affiliation{ Department of Chemistry,\\ Indian Institute of Technology Bombay,\\ Powai, Mumbai, PIN: 400076, India}
\author{Rahul Maitra}
\email{rmaitra@chem.iitb.ac.in}
\affiliation{ Department of Chemistry,\\ Indian Institute of Technology Bombay,\\ Powai, Mumbai, PIN: 400076, India}

\title{A Synergistic Approach towards Optimization of Coupled Cluster Amplitudes by Exploiting Dynamical Hierarchy}
%\linenumbersCorrections

\begin{abstract}

The coupled cluster iteration scheme for determining the
cluster amplitudes involves a set of nonlinearly 
coupled difference equations. In the space spanned by the 
amplitudes, the set of equations are analysed as a
multivariate time-discrete map where the concept of time
appears in an implicit manner. With the observation that
the cluster amplitudes have difference in their
relaxation timescales with respect to the distributions
of their magnitudes, the coupled cluster iteration 
dynamics are considered as a synergistic motion of
coexisting slow and fast relaxing modes, manifesting a
dynamical hierarchical structure. With the identification
of the highly damped auxiliary amplitudes, their time
variation can be neglected compared to the principal
amplitudes which take much longer time to reach the fixed 
points. We analytically establish the adiabatic 
approximation where each of these auxiliary amplitudes
are expressed as unique parametric functions of the 
collective principal amplitudes, allowing us to study the 
optimization with the latter taken as the independent 
degrees of freedom. Such decoupling of the amplitudes 
significantly reduces the 
computational scaling without sacrificing the accuracy
in the ground state energy as demonstrated by a number of
challenging molecular applications. A road-map to treat
higher order post-adiabatic effects is also discussed.
\end{abstract}

\maketitle

\section{Introduction}
Coupled Cluster Theory (CC)\cite{cc3,cc4,cc5,bartlett2007coupled,crawford2000introduction} is well established 
electronic structure methodology for accurately solving molecular energetics and properties. In CC method,
a correlated wavefunction is generated by the action of 
an exponential wave-operator involving rank-one ($T_1$), rank-two ($T_2$), ..., rank-$n$ ($T_n$) cluster operators, that act 
on a suitably chosen reference determinant: $|\psi_{CC}\rangle = e^{T_1 + T_2 + T_3+ ...}|\phi_0\rangle$. 
For closed shell cases, the reference determinant $\phi_0$ is often taken as the Hartree-Fock determinant and
the amplitudes ($t$) associated with the cluster operators are the unknown quantities 
that are self-consistently optimized. The cluster amplitudes associated with $T_k$ are determined by 
projecting a similarity 
transformed effective hamiltonian $\hat{H}_{eff} = e^{-T} \hat{H} e^T$ against the $k-$tuply excited determinant
that, in principle, folds in the effects of excited determinants such that the correlated 
ground state energy can be determined as an expectation value of $\hat{H}_{eff}$ with respect to the 
reference determinant.

CC theory is size-extensive and size-consistent at any level of truncation in the rank of 
the cluster operators. In this manuscript, without any loss of generality, we would 
consider the case where only rank-two cluster operators are taken. The resulting CC theory 
with doubles (CCD) involves coupled optimization of $n_o^2 n_v^2$ amplitudes, where $n_o$
and $n_v$ are the number of occupied (hole, to be denoted as $i,j,k,...$) and unoccupied 
(virtual, denoted as $a,b,c,...$) orbitals. Due to the 
exponential structure of the waveoperator, any optimization strategy involves nonlinearly 
coupled set of equations among $n_o^2 n_v^2$ amplitudes. This is often done by iterative 
minimization of a residue vector $R_{ijab}= \bra{\phi_{ij}^{ab}} e^{-T} \hat{H} e^T \ket{\phi_0}$,
where $\ket{\phi_{ij}^{ab}}$ is the excited determinant. Thus, at the \textit{fixed point} of
the nonlinear optimization process, as $R_{ij}^{ab} \rightarrow 0, \forall i, j, a, b$, 
$T \rightarrow T_{CCD}$. Note that due to the excitation structure of the cluster operators,
the cluster operators are only allowed to contract with the hamiltonian and thus, this
optimization procedure scales as $n_o^2 n_v^4$ at the worst.

Due to the nonlinear structure of the working equations, 
it is alluring to visualize the iterative scheme as 
a time-discrete multivariate map where \enquote{time} enters in
an implicit manner with each iterative 
step being embedded as one discrete-time step.
Such iteration scheme undergoes chaotic dynamics when the system of equations 
is perturbed with an input perturbation. The stability of the equations were, for the first time, demonstrated by 
Szak{\'a}cs and Surj{\'a}n\cite{szakacs2008stability,szakacs2008iterative} which is later extended to 
Lippmann-Schwinger equation\cite{surjan2022stability}. Some of the present authors
led the concept one step further to demonstrate that under the influence of single source of 
perturbation, the nonlinear iteration scheme essentially
behaves like a multivariate time-discrete map of one-parameter family\cite{agarawal2020stability} which obeys the universality of the
Feigenbaum dynamics\cite{feigenbaum1978quantitative,feigenbaum1979universal}. The authors argued the
existence of a set of \textit{dominant} collective modes that macroscopically govern the optimization 
process. Contrarily, there exists large number of other \textit{recessive} variables (amplitudes)
that essentially evolve synergistically as dictated by the dominant ones. The present 
authors, for the first time, exploited the synchronization among the cluster 
amplitudes during the optimization trajectory via machine learning that resulted tremendous reduction 
in the degrees of freedom and $40\%-50\%$ savings in computational 
time to achieve sub-microHartree ($\mu E_h$) \cite{agarawal2021accelerating, agarawal2022hybrid}. As a matter of fact, 
such a synchronous evolution of all the variables during a (discrete) time evolution is a
central theme of Synergetics which gives us a prescription to write the entire dynamics through 
the collective dominant modes only. Here we briefly explain the concept of nonlinear dynamics and synergetics
and how one may apply the concepts of dimensionality reduction for the CC optimization strategy.

In the regions near the fixed point 
equilibrium, 
the linear stability analysis enables us to classify the 
variables into two different sets: the 
auxiliary 
(recessive) and principal (dominant) 
modes\cite{Haken_1989, haken1982slaving}. 
The auxiliary modes are the
ones that relax much faster than the principal 
modes\cite{wunderlin1987slaving}, 
and in
the characteristic timescale of the principal 
modes, one can
neglect the time variation of the auxiliary 
modes (known as Adiabatic Approximation)\cite{Haken1983}. This 
is 
conceptually the generalised version of the Born-Oppenheimer
approximation\cite{born1927born}. For most of the practical cases 
the following conditions 
hold\cite{synergetics1976introduction}:
\begin{enumerate}
\item the number of principal 
modes are much less than than the number of auxiliary modes
\item the magnitude of the principal  modes are significantly 
larger than the magnitude of the auxiliary modes. 
\end{enumerate}
Given that the system of variables can be grouped into the auxiliary and 
principal modes, the \enquote{Slaving Principle} 
enables us to express each of the auxiliary modes as unique
function of the principal modes. This thus allows us to confine
our attention only to the dynamics of the principal modes. 
The auxiliary modes unanimously follow the \enquote{orders} 
of the principal modes and therefore, this is known as the 
master-slave dynamics. 
With the time evolution of the principal modes, the auxiliary 
modes also get updated according to their parametric 
dependence on the principal modes. However,
there exists \textit{circular causal} relationship such 
that the updated information of the auxiliary modes gets 
coupled back to the dynamical equations of the principal modes,
which may be referred to as the feedback 
coupling\cite{synergetics1976introduction}. 
The whole dynamical system thus
evolves in the interdependent way. As the auxiliary modes at 
each step merely chip in as a parametric function
of the principal modes, the effective degrees of freedom of 
the whole system is significantly reduced. 
In one of our previous publications, we developed a
first-principle based CC optimization strategy, termed as 
the Adiabatically Decoupled Coupled Cluster (ADCC),
where we have shown that it is possible to distinguish the 
various amplitudes at their MP2 level into principal and auxiliary 
modes. Based on certain assumptions which
are conceptually close in spirit to the adiabatic 
approximation, the authors were able to reproduce very 
accurate results by optimizing the variables in a reduced 
subspace\cite{agarawal2021approximate}. 

The main purpose of this article is to show that some of these
features of nonlinear dynamics for the systems close to classical critical points can also be applied in numerically accurate manner (although not mathematically 
exact) to reduce the 
computational scaling where nonlinear iterative optimization is involved. 
We will also demonstrate how the \enquote{adiabatic 
approximation} comes out quite naturally in a more general and 
rigorous mathematical way, given that some of the cluster amplitudes
have significantly longer time scale of relaxation to converge. 
This will also elucidate the scope to 
further improve the ADCC results by incorporating some of the 
higher order post-adiabatic terms\cite{haken1975higher,wunderlin1981generalized}. Towards this, we will first 
present mathematical preliminaries, taking general class of 
dynamical systems that motivate us towards
the development of the adiabatically decoupled CC iterative
formulation. We will mainly focus on the underlying 
concepts of master and slave variables, which are related by 
the \textit{slaving principle}, and will show how the CC 
iterative scheme can be adapted within the master-slave 
dynamics framework for dimensionality reduction. We will show the 
performance of our model as a function of the dimension of
the master modes to justify that CC theory in principle can be
optimized in significantly reduced dimension than that 
dictated by the size of the basis set.

\section{Adiabatically Decoupled Scheme for Coupled Cluster Amplitude Optimization:}

\subsection{General Mathematical Preliminaries towards
Dimensionality Reduction: Concepts of Adiabatic Approximation
and Slaving Principle}

%There exist a specific kind of multi-variable dynamical systems in nature where the huge number of variables can be 
%sub-divided into two groups - auxiliary and unauxiliary modes. Due to their larger damping the auxiliary modes are those that reach the 
%equilibrium or the fixed point much faster than the unauxiliary modes. According to the Slaving Principle in Synergetics - in the 
%characteristic time scale of the unauxiliary modes, the time variation of the auxiliary modes can be neglected. With some further 
%approximations we can finally express the auxiliary modes as a function of unauxiliary modes. Generally for all practical cases the number 
%of auxiliary modes in a system is significantly higher than the number of unauxiliary modes and thus the functional relationship between 
%the two hugely reduce the effective degrees of freedom of the dynamics.

As mentioned in the introduction, for specific multivariate 
dynamical systems, one may divide the entire variable space
into auxiliary and principal modes having large difference in
their magnitude and characteristic timescale of relaxation.
In order to have an easy readability, in the following, we 
first discuss the mathematical 
preliminaries that allow us to express the dynamics of the
whole system solely in terms of the principal modes. We will
consider a general dynamical system with continuous time
variation. The corresponding discrete analogue pertinent
to the CC iterative scheme will be developed in the next 
subsection. 

Let us consider a multivariate dynamical system 
$\{ \vectorbold{\hat{q}} \}$ evolving according to the 
following general form of an equation:
\begin{equation}
    \vectorbold{\dot{\underline{\hat{q}}}} = L \vectorbold{ \underline{\hat{q}}} + \vectorbold{N} (\{ \vectorbold{\hat{q}} \})
\end{equation}
where, $L$ is a coefficient matrix independent of the
dynamical variables and $\vectorbold{N}$ contains all the 
nonlinear couplings among them. One may perform a linear
stability analysis\cite{strogatz2018nonlinear,alligood1996two,lam2003introduction} to obtain 
a set of eigenvalues $\{\lambda\}$ of the stability matrix that can be classified 
into the following two categories:\cite{wunderlin1981generalized, synergetics1976introduction}
\begin{equation} 
\begin{split}
  & \{ {\lambda}_u \} \gtrless  0 ; \hspace{5mm} \mbox{(total number }N_u) \\
  & \{ {\lambda}_s \} < 0;      \hspace{6mm} \mbox{(total number } N_s)
\end{split}
\label{lambda_values}
\end{equation}
with the condition: 

\begin{equation} \label{lambda_condition}
    \mid \lambda_{s_i}\mid > \mid \lambda_{u_I} \mid    
\end{equation}

 In the usual scenario, $N_s >> N_u$, where the individual elements from the 
 sets of eigenvectors of $\lambda_u$ ($\{u\}$) 
and $\lambda_s$ ($\{s\}$) are denoted by $u_I$ and $s_i$. 
Positive values of $\lambda_u$ indicates the dynamics is non-equilibrium and the corresponding eigenvectors are the unstable modes which move away from the fixed point as time progresses\cite{alligood1996two}. This justifies the subscript $u$ and $s$ referring to unstable and stable modes, respectively. On the other hand, keeping in mind the CC iteration dynamics, we are only interested here in the negative $\lambda_u$ cases, referring to an overall equilibrium system that eventually reaches one of the fixed points of the dynamics. This indicates that all the $\lambda$s can be considered as damping factors. Hence, we drop the terminology of unstable and stable modes from here on, rather we keep the same subscripts but prefer to refer them in a more general way as the \textit{principal} and \textit{auxiliary} modes. In the transient period, the variables with smaller and larger amplitudes generally have larger and smaller damping factors respectively, allowing us to establish a demarcation between auxiliary and principal amplitudes. The nonlinearly coupled 
equations of motion for these two sets of amplitudes 
can be written as:
\begin{equation} \label{2D_udot}
    \dot{u_I} = \lambda_{u_I} u_I + Q_I({u,s}); \hspace{5mm} \mbox{where } I=1,2,...,N_u
\end{equation}
and
\begin{equation}\label{2D_sdot}
    \dot{s_i} = \lambda_{s_i} s_i + P_i({u,s}); \hspace{5mm} \mbox{where } i=1,2,..,N_s
\end{equation}
%before the equation
%\vspace{10mm}
%\textcolor{blue}{ The positive or negative sign of $\lambda_u$ corresponds to non-equilibrium or equilibrium systems respectively.  The corresponding
%eigenvectors of $\lambda_u$ and $\lambda_s$ form the principal and auxiliary modes. However, in our case we are only interested in equilibrium systems where we classify auxiliary and principal modes based on their 
%relaxation timescale in the transient period of the system.}
Here $Q_I$, $P_i$ contains all the information of nonlinearity and 
inter-mode 
coupling. In a narrow region around a fixed point 
(i.e. $\{\dot{u_I}\},\{\dot{s_i}\}=0$) where linearization can be applied, 
one may analyze the dynamics by neglecting the small contributions from
the nonlinear terms in $Q_I$ and $P_i$. In such a scenario, when the system is 
perturbed slightly from the fixed point equilibrium, the whole system moves 
under the influence of the principal modes $\{u_I\}$ in their characteristic timescale,
whereas, the auxiliary modes $\{s_i\}$ decay back to the fixed point.
In other words, $s_i$ can be considered as heavily damped modes
whereas $u_I$ is relatively under-damped.

By exploiting the characteristic relaxation timescale,
we wish to show that $s_i$ amplitudes can 
be expressed by means of $u_I$ only, such that
one can also eliminate $\{s_i\}$ from Eq. \eqref{2D_udot} by direct substitution. 
The most general solution to Eq. \eqref{2D_sdot} 
can be obtained as
\begin{equation} 
\begin{split}
  & \Big( \frac{d}{dt} - \lambda_{s_i} \Big) s_i = P({u,s}) \\
\implies   &  s_i = \Big( \frac{d}{dt} - \lambda_{s_i} \Big)^{-1} P(\{u,s\}) \\ 
\end{split}
\end{equation}
The inverse differential operator can be solved with an 
integral form that allows us to write\cite{wunderlin1981generalized}: 
\begin{equation} \label{s2D_int}
    s_i(t) = \int_{-\infty}^{t} e^{\lambda_{s_i} (t-\tau)}  P(\{u,s\}) d{\tau}
\end{equation}
Assuming the integral is finite and well-behaved\cite{schmidt1985haken, haken1983nonlinear}, one may perform an integration by parts to
further simplify Eq. \eqref{s2D_int}
\begin{equation} \label{s2D}
    s_i(t) = -\frac{1}{\lambda_{s_i}} P({\{u,s\}}) - \frac{1}{(-\lambda_{s_i})}\int_{-\infty}^{t} e^{\lambda_{s_i} (t-\tau)} (\frac{d}{d\tau} P) d{\tau}
\end{equation}
The second term in right hand side onward are usually very small. Hence, neglecting the 
small higher order terms, Eq. \eqref{s2D} reads
\begin{equation} \label{s}
    s_i(t) = -\frac{1}{\lambda_{s_i}} P(\{u,s\})
\end{equation}
A careful observation leads to the interesting fact 
that we can avoid all the mathematical jargon and 
still get the same result as that of Eq. \eqref{s} 
if we simply set $\dot{s_i}=0$ in Eq. \eqref{2D_sdot}. 
This simply implies that in the characteristic time scale 
of the principal modes, the time-variation of the auxiliary 
modes can be neglected. This is to be referred as the 
\textit{adiabatic approximation}.

Generally in all practical cases, the amplitude of the principal modes are extremely 
large in comparison to auxiliary modes. Thus all the contributions from $s_i$ can 
be neglected compared to $u_I$ such that the right hand side of Eq. \eqref{s} can be written 
as a function of $u_I$ alone, which simplifies the expression as: 
\begin{equation} \label{s_ad}
    s_i(t) = -\frac{1}{\lambda_{s_i}} P(\{u\})
\end{equation}
As we have the expression in Eq. \eqref{s_ad} for the auxiliary modes as a function of the principal
modes, we can now imitate the system dynamics in a reduced subspace spanned only by the 
principal modes, which are much fewer in number compared to the auxiliary modes. In the context
of CC iterative approach, our goal is to accurately determine $s$ as a function of $\{u\}$ only,
which would be fed back to the dynamics of $\{u\}$ to get an updated set of principal modes.
This loop, referred to as the \textit{circular causality loop}, would continue till the 
principal modes converge below a pre-defined threshold.

\subsection{Coupled Cluster iterative optimization from the synergistic viewpoint:}
We now turn our attention to the optimization of CC theory using Jacobi iterative scheme. We would
consider the iterative scheme as a time-discrete map where each iteration is embedded as
one time step. Furthermore, since the iterative optimization leads to a set of fixed points,
it can be considered as an equilibrium system from the perspective of nonlinear science, and hence 
the demarcation of the auxiliary and principal modes (as shown in Eq. \eqref{lambda_values} for
non-equilibrium systems) does not strictly hold according to the linear stability 
analysis\cite{agarawal2020stability}. In general, for equilibrium systems (in the perspective of
nonlinear dynamics), all the values of $\lambda$ may be negative; however, for all practical 
purposes, one may bypass the linear stability criteria and 
classify the amplitudes based on their distribution of magnitudes into auxiliary and principal
modes having shorter or longer characteristic relaxation time to reach their fixed point 
solutions. We will generally refer to these amplitudes as auxiliary and principal amplitudes,
respectively. 

With the approximations in mind, let us now consider the amplitudes updating Jacobi equation
\begin{equation} \label{eq1}
\begin{split}
 \Delta t_{\mu} & = t_{\mu}^{(k+1)} - t_{\mu}^k = \frac{R_\mu}{D_\mu}\\
 & =\frac{1}{D_\mu} (H_\mu + (\contraction{}{H}{}{T} H T)_\mu + \frac{1}{2}(\contraction{}{H}{}{T}
\contraction[2ex]{}{H\;}{T}{}
H\;TT)_\mu ) 
\end{split}
\end{equation}
Here $T$ is the cluster operator, $t_\mu$ denotes the
associated amplitudes with hole-particle excitation
structure \enquote*{$\mu$} and $R_\mu$ is the corresponding residue. 
The hamiltonian $H$ contains the one and two
electron terms. It was observed from a number of numerical examples\cite{agarawal2021approximate}
that the amplitudes with large magnitude (at the MP2 level) take substantially more 
number of iterations to reach their converged values (fixed points) than the ones 
with smaller magnitudes. Based on the relative magnitude of these amplitudes, the entire 
amplitude space can be subdivided into a Large Amplitude Subset (\textit{LAS}, spanned by 
$\{T_L\}$ with dimension $n_L$) and Small Amplitude Subset (\textit{SAS}, spanned by $\{T_S\}$, with 
dimension $n_S$). In our time-discrete iterative map, their amplitudes $\{t_L\}$ plays the 
role of the principal modes and $\{t_S\}$ plays the role of the auxiliary modes. Note also that $n_S >> n_L$ and $\mid{t_L}\mid > \mid{t_S}\mid$. They will thus be referred 
to as the \textit{principal} and \textit{auxiliary} amplitudes, respectively. In conjunction to
the dynamics of the principal and auxiliary modes shown in Eqs. \eqref{2D_udot} and \eqref{2D_sdot}, 
the same for the principal and auxiliary amplitudes can be written as:
\begin{equation} \label{del_tL}
\begin{split}
    \Delta t_{L_I} &= \frac{1}{D_{L_I}}{(\contraction{}{H^d_{L_I}}{}{T_{L_I}} H^d_{L_I} T_{L_I})}_{L_I} + Q_{L_I}(\{T_L,T_S\}); \forall I= 1, 2, ..., n_L
\end{split}
\end{equation}
and
\begin{equation} \label{del_tS}
\begin{split}
    \Delta t_{S_i} &= \frac{1}{D_{S_i}}{(\contraction{}{H^d_{S_i}}{}{T_{S_i}} H^d_{S_i} T_{S_i})}_{S_i} + P_{S_i}(\{T_L,T_S\}); \forall i= 1, 2, ..., n_S
\end{split}
\end{equation}
where, $L_I$ and $S_i$ are composite hole-particle indices 
for \textit{LAS} and \textit{SAS} elements, respectively.
$D_{\mu}$ is the corresponding orbital energy difference 
for the general hole-particle index $\mu$ . 
Here $Q_{L_I}$ and $P_{S_i}$ contain all the information of nonlinearity for the 
dynamics of the principal and auxiliary amplitudes, and they can be expanded as:
\begin{equation} \label{Q}
\begin{split}
& Q_{L_I}(\{T_L,T_S\}) = \frac{1}{D_{L_I}} \Big[ \Big(H_{L_I} +
{{(\contraction{}{H}{}{T_{L_J}} H T_{L_J})}_{L_I}} + {{(\contraction{}{H}{}{T_{S_i}} H T_{S_i})}_{L_I}}\Big) + \\
& \frac{1}{2} \Big({(\contraction{}{H}{}{T_{L_J}} \contraction[2ex]{}{H\;}{T_{L_K}}{} H\; T_{L_J} T_{L_K})_{L_I}}
+ {(\contraction{}{H}{}{T_{L_J}} \contraction[2ex]{}{H\;}{T_{S_i}}{} H\; T_{L_J} T_{S_i})_{L_I}}+{(\contraction{}{H}{}{T_{S_i}} \contraction[2ex]{}{H\;}{T_{S_j}}{} H\; T_{S_i} T_{S_j})_{L_I}})\Big) \Big] 
\end{split}
\end{equation}
and
\begin{equation}\label{P}
\begin{split}
& P_{S_i}(\{T_L,T_S\}) = \frac{1}{D_{S_i}} \Big[ \Big(H_{S_i} +
{{(\contraction{}{H}{}{T_{L_I}} H T_{L_I})}_{S_i}} +
{{(\contraction{}{H}{}{T_{S_j}} H T_{S_j})}_{S_i}}\Big) + \\ 
& \frac{1}{2} \Big({(\contraction{}{H}{}{T_{L_I}} \contraction[2ex]{}{H\;}{T_{L_J}}{} H\; T_{L_I} T_{L_J})_{S_i}}
+{(\contraction{}{H}{}{T_{L_I}} \contraction[2ex]{}{H\;}{T_{S_j}}{} H\; T_{L_I} T_{S_j})_{S_i}}+{(\contraction{}{H}{}{T_{S_j}} \contraction[2ex]{}{H\;}{T_{S_k}}{} H\; T_{S_j} T_{S_k})_{S_i}})\Big) \Big] 
\end{split}
\end{equation}

\begin{figure*}[!ht]
    \centering
\includegraphics[angle=00,scale=1.0]{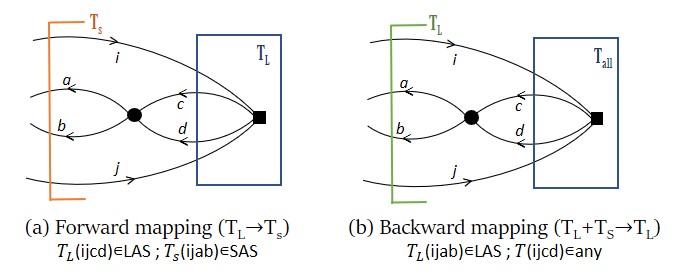}
\caption{Representative diagrams contributing to the leading
order of computational scaling. Diagrams of the same topology are shown for forward mapping (a) and backward mapping (b) with different interpretation.}
    \label{fig:adcc_diag}
\end{figure*}

Here the \textit{diagonal} terms, $H^d$, in Eqs. 
\eqref{del_tL} and \eqref{del_tS} needs some explanation. 
The diagonal part of the hamiltonian matrix $H^d_{\mu}$
are the composite tensorial terms which upon contraction
with $T_{\mu}$ generate a structure with hole-particle 
composite structure as $\mu$. The diagonal part is extracted 
out in the expressions of $\Delta t_\mu$ in Eqs. 
\eqref{del_tL} and \eqref{del_tS} to show their one-to-one
correspondence with Eqs. \eqref{2D_udot} and \eqref{2D_sdot}.
The explicit expression for $H^d_{\mu}$ is given by    
\begin{multline}
    H_\mu^d = (1 + P(i,j)P(a,b)) \Big(-f_{ii}+f_{aa}+\frac{1}{2}v_{ab}^{ab}+2v_{ia}^{ai}\\-(1+\delta_{ij}\delta_{ab}) v_{ia}^{ia}+ \frac{1}{2}v_{ij}^{ij}- v_{ib}^{ib} \Big)\\
    \mu = \{ijab\} \in \mbox{\textit{SAS} or \textit{LAS}}
    \label{eqhd}
\end{multline}
Here $i,j$ refer to the occupied (hole) orbital 
indices while $a,b$ refer to the unoccupied (particle) 
orbital indices. The canonical Fock operator matrix 
elements and the two electron integrals are represented 
by \textit{f} and \textit{v} respectively, and $P$ is the permutation operator.

We are now in a position to develop the discrete-time 
analogue of Eq. \eqref{s2D_int} for the CC optimization. 
Without going into the details of the equation, for which 
we refer to Haken\cite{haken1997discrete,haken1982slaving}, the most general solution for the
auxiliary amplitudes, $t_S$, can be written as:
\begin{equation}
    t_{S_i} = \sum_{m=-\infty}^{l} (1+ \lambda_{S_i})^{l-m} P_{S_i}
\end{equation}
where, 
\begin{equation} \label{lambda}
    {\lambda_{S_i} = \frac{H^d_{S_i}}{D_{S_i}}}
\end{equation}
Note that $H^d_{S_i}$ and $D_{S_i}$ have opposite sign (and
hence CC equations form a convergent series) and thus 
$\lambda_{S_i}$ is negative. The expression in the right
hand side of Eq. \eqref{lambda} can be expanded using 
\textit{summation by parts}, a mathematical trick similar to
integration by parts of Eq. \eqref{s2D}, to obtain:
\begin{equation} \label{general_ts} 
\begin{split}
  & t_{S_i} = P_{S_i} \sum_{m=-\infty}^{l} (1+ \lambda_{S_i})^{l-m} \\ 
  & - \sum_{m=-\infty}^{l} (1+ \lambda_{S_i})^{l+1-m} \sum_{m'=-\infty}^{m-1} (1+ \lambda_{S_i})^{m-1-m'} \Delta P_{S_i} 
\end{split}
\end{equation}
Here, \enquote{$\Delta$} is the time-discrete version of the
time derivative operator \enquote{$\sfrac{d}{dt}$}. Since the 
auxiliary modes have $\lambda_{S_i} < 0, \forall S_i$, one 
can simplify the intricate summations above using the 
identity: 
\begin{equation}
    \sum_{m=-\infty}^{l} (1+ x)^{l-m} = -\frac{1}{x} ; \hspace{1mm} \mid x+1 \mid <1
\end{equation}
The simplification of Eq. \eqref{general_ts} leads to
\begin{equation} \label{ts_ad+p_ad}
    t_{S_i} = \underbracket[0.8pt]{-\frac{P_{S_i}(\{t_L,t_S\})}{\lambda_{S_i}}}_\text{\clap{adiabatic}}  \overbracket[0.8pt]{-\frac{\Delta P_{S_i}(\{t_L,t_S\})}{\lambda_{S_i}^2}}^\text{\clap{post-adiabatic}}    
\end{equation}
We will briefly explain the significance of these two terms
that appear on the right hand side of Eq. \eqref{ts_ad+p_ad}.
As mentioned previously, one may neglect the time variation 
of the auxiliary modes (auxiliary amplitudes) in the 
characteristic timescale of the principal modes (principal
amplitudes). By setting up the condition 
$\Delta{t_{S_i}} = 0$ to neglect the time variation of the
auxiliary amplitudes, one obtains the first term on the
right hand side of Eq. \eqref{ts_ad+p_ad}. This thus can
be interpreted as the \textit{adiabatic approximation}. Note
that in one of our earlier publications\cite{agarawal2021approximate}, the adiabatic 
approximation was introduced simply by setting the 
decoupling condition. Here we further corroborate the concept
by introducing a discrete-time dependent picture and 
timescale decoupling in the amplitude space. 
The second term on the right hand side of 
Eq. \eqref{ts_ad+p_ad} depends on the discrete-time variation
of the (part of the) residue $P_{S_i}$ and depends on 
$\lambda_{S_i}^2$. This thus may be interpreted as post-
adiabatic correction and will not be considered in this
article any further. We will introduce the post-adiabatically
corrected optimization scheme in our future publication.

In what follows, we will now proceed exactly the same way 
that allowed us to derive Eq. \eqref{s_ad} from 
Eq. \eqref{s}. Noting the fact\cite{agarawal2021approximate}
the auxiliary amplitudes $t_S$ are significantly small
compared to the principal amplitudes, $t_L$, one may
neglect $t_S$ from the adiabatically decoupled expression 
of $t_{S_i}$:
\begin{equation} \label{ts_adiab}
    t_{S_i(ad)} = -\frac{P_{s_i}(\{t_S,t_L\})}{\lambda_{S_i}} \xrightarrow{{\mid t_S \mid } \approx 0} -\frac{P_{s_i}(\{t_L\})}{\lambda_{S_i}}
\end{equation} 
or in the long hand notation by writing $P_{S_i}(\{t_L\})$ explicitly:
\begin{equation} \label{ts_long_hand}
    t_{S_i(ad)} = - \frac{1}{H^d_{S_i}} \Big[ H_{S_i} + {{(\contraction{}{H}{}{T_{L_I}} H T_{L_I})}_{S_i}} + \frac{1}{2} \Big({(\contraction{}{H}{}{T_{L_I}} \contraction[2ex]{}{H\;}{T_{L_J}}{} H\; T_{L_I} T_{L_J})_{S_i}} \Big)    \Big] 
\end{equation}
where all the terms that involve $T_S$ are set to zero.
We note that setting the adiabatic condition 
$\Delta{t_{S_i}} = 0$ \textit{does not} imply that the
auxiliary amplitudes are frozen to given pre-computed values.
Rather, they are updated through parametric functional
dependence on the principal amplitudes. We now turn our 
attention to the computation of the determination of the
principal amplitudes via feedback coupling.

\begin{figure*}[!ht]
    \centering
\includegraphics[width=\textwidth]{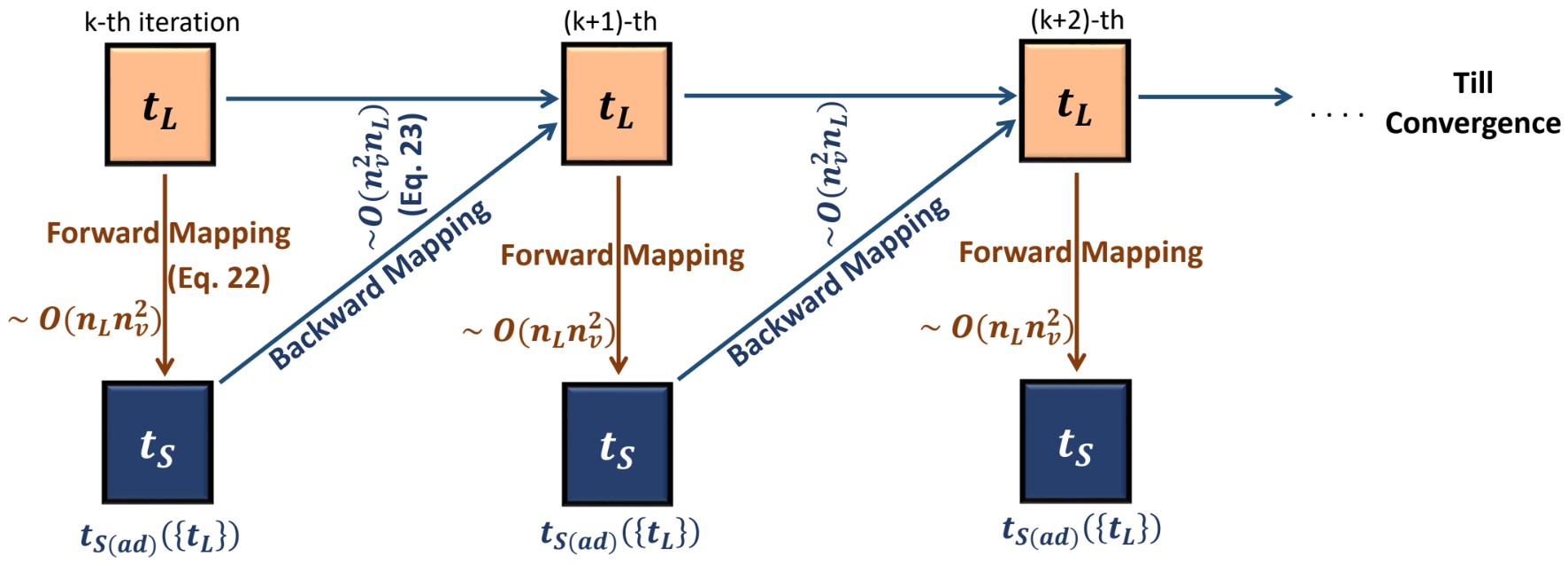}
\caption{A schematic representation of the circular algorithm where the auxiliary amplitudes, $t_S$ are determined through sole function of the principal amplitudes, $t_L$, which are in turn
updated through the back coupling of both sets of amplitudes. At each of these steps, the computational scaling goes as $n_Ln_v^2 << n_o^2n_v^4$.}
    \label{fig:iteration}
\end{figure*}

\subsubsection*{Feedback coupling and the determination of the principal amplitudes:}

As previously mentioned, there exists circular causality 
relationship among the principal and the auxiliary amplitudes.
This implied that the updated information of the auxiliary
amplitudes (obtained from Eq. \eqref{ts_long_hand}) must
get fed back to the updating equations for $T_L$. Noting the
fact that $T_{s_i(ad)} = T_{S_i}(\{T_L\})$, one may explicitly
write the principal amplitude update equations as:
 \begin{equation} \label{RL}
\begin{split}
& \Delta t_{L_I} = \frac{1}{D_{L_I}}\Big[ \Big(H_{L_I} +
{{(\contraction{}{H}{}{T_{L_J}} H T_{L_J})}_{L_I}} + {{(\contraction{}{H}{}{T_{S_i(ad)}} H T_{S_i(ad)})}_{L_I}}\Big) + \\
& \frac{1}{2} \Big({(\contraction{}{H}{}{T_{L_J}} \contraction[2ex]{}{H\;}{T_{L_K}}{} H\; T_{L_J} T_{L_K})_{L_I}}
+ {(\contraction{}{H}{}{T_{L_J}} \contraction[2ex]{}{H\;}{T_{S_i(ad)}}{} H\; T_{L_J} T_{S_i(ad)})_{L_I}}+ \\
& {(\contraction{}{H}{}{T_{S_i(ad)}} \contraction[2ex]{}{H\;}{T_{S_j(ad)}}{} H\; T_{S_i(ad)} T_{S_j(ad)})_{L_I}})\Big) \Big] 
\end{split}
\end{equation}
which can be written as:
 \begin{equation} \label{RL1}
\begin{split}
& \Delta t_{L_I} = \frac{1}{D_{L_I}} \Big(H_{L_I} +
{{(\contraction{}{H}{}{T} H T)}_{L_I}} + \frac{1}{2} {(\contraction{}{H}{}{T} \contraction[2ex]{}{H\;}{T}{} H\; T T)_{L_I}}\Big) 
\end{split}
\end{equation}
where 
\begin{equation}
T=\sum_I T_{L_I} \oplus \sum_i T_{S_i(ad)}(\{T_L\})    
\end{equation}
This implies that both the principal and auxiliary amplitudes
couple to the equations of the former. Since the contribution 
of the auxiliary amplitudes is significantly smaller, one may
truncate the right hand side of Eq. \eqref{RL} by keeping
only those terms containing $T_S$ which are overall linear. 
Depending on which terms are retained, we have developed two
schemes. In scheme-I, all the terms in the right hand side of
Eq. \eqref{RL} are retained. This implies that scheme-I 
takes care of complete feedback coupling. In scheme-II, only
terms I-IV in the right hand side of Eq. \eqref{RL} are
retained. In other words, the feedback of the auxiliary 
amplitudes in the equation of the principal amplitudes is
taken up to overall linear order.

\subsubsection*{Analysis of computational scaling:}

\begin{figure*}[!ht]
    \centering  
\includegraphics[width=\textwidth]{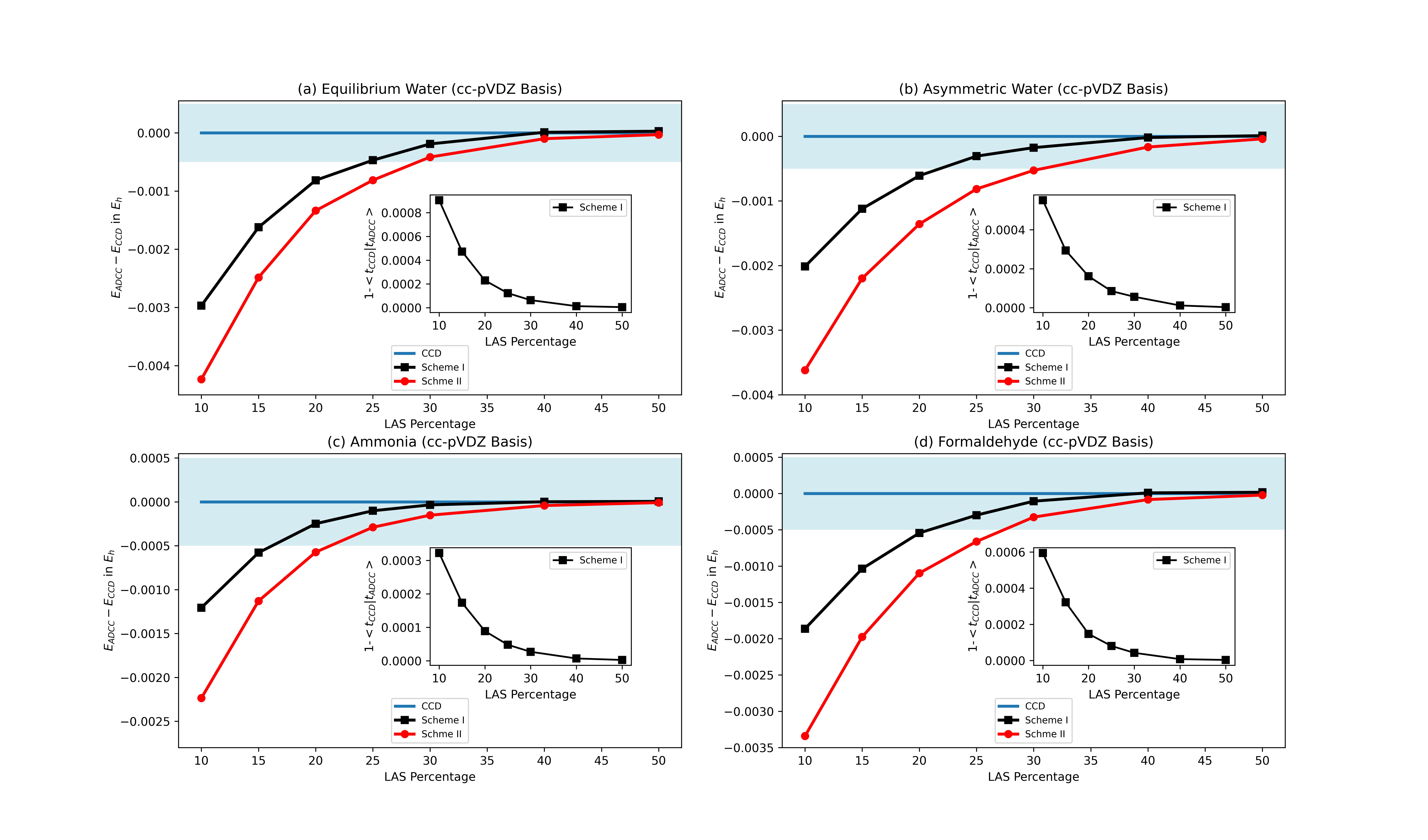}
\caption{Comparative study of the accuracy of ADCCD scheme-I (black line) and scheme-II (red line) against canonical CCD as functions of \textit{LAS} size for four different molecules. Note that the horizontal blue line denotes the canonical CCD energy scaled to zero and the blue shaded region denotes $\pm0.5 mE_h$ region. ADCCD scheme-I is slightly more accurate over scheme-II, particularly with smaller \textit{LAS}, as expected. ADCCD scheme-I achieves $\sim 0.01 mE_h$ accuracy with only 
about 30 \% of the total nonzero amplitudes taken as independent (principal) ones. The inset shows the accuracy of the combined set of cluster amplitudes as compared to canonical CCD and the former is shown to be as accurate as the latter with LAS taken to be 40$\%$ where one obtains accuracy up to $\mu E_h$ precision.}
    \label{fig:adcc_ccd_plots}
\end{figure*}

We now briefly discuss about the computational scaling 
associated with the determination of the auxiliary 
amplitudes (Eq. \eqref{ts_long_hand}) and principal
amplitudes (Eq. \eqref{RL}). For detailed analysis of the
computational scaling, we refer to one of our earlier 
publications\cite{agarawal2021approximate}. Here, for our
analysis of the scaling, we would only consider the most
expensive linear diagram that appears in CC theory, namely the
one with all particle contraction. Such a representative 
diagram is shown in Fig. \ref{fig:adcc_diag}. 
Fig. \ref{fig:adcc_diag}(a) represents its 
interpretation in the context of the forward mapping via 
Eq. \eqref{ts_long_hand} where $t_L$ amplitudes uniquely
determine the $t_S$ amplitudes. Diagrammatically, 
\textit{only} the $T_L$ operators (filled squared box, 
of dimension $n_L$) contracts with the hamiltonian 
(filled circles) to generate $T_S$. Here the index quartet 
of $T_L$, $(ijcd)$, necessarily belongs to one of the $n_L$
elements of \textit{LAS}. On the other hand, the uncontracted 
particle indices $a,b$ can be arbitrary with the constraint 
that the uncontracted index quartet $(ijab)$ should necessarily
belong to one of the \textit{SAS} elements. This automatically
renders the leading computational scaling for the 
forward mapping to be $n_Ln_v^2$. 

We now turn our attention to the scaling of the backward 
mapping (Eq. \eqref{RL}). For this purpose, we resort to
the diagram of the same topology as we discussed for the 
forward mapping, and interpret it in a different way as 
shown in Fig. \ref{fig:adcc_diag}(b). In this case,
\textit{all} the amplitudes are allowed to couple with
the hamiltonian, however, the resulting structure should 
have the corresponding amplitude in \textit{LAS}. This implies that
the index quartet $(ijcd)$ can be any one of the 
principal or auxiliary amplitudes. However, the uncontracted
index quartet $(ijab)$ should necessarily belong to one
of the $n_L$ elements of $T_L$. The restriction on the
uncontracted index to the ones belonging to \textit{LAS} automatically
ensures that $i$ and $j$ cannot take any arbitrary hole
orbital level. The scaling for the construction of this
diagram is again $n_Ln_v^2$ at the worst.

Thus the overall scaling of the scheme never exceeds 
$n_Ln_v^2$ for the ADCC scheme, whereas, for CCD, the same
scales as $n_o^2n_v^4$. Since $n_L$ is only a small fraction
(vide infra) compared to $n_o^2n_v^2$, there is a clear
computational advantage of our scheme offers over the 
conventional one.

In summary, the overall algorithm moves in a circular manner 
where at each step, the auxiliary amplitudes are determined 
via Eq. \eqref{ts_long_hand} as functions of $\{t_L\}$ alone. 
On the other hand, the principal amplitudes, $t_L$ are 
updated through complete or partial feedback coupling (for 
scheme-I and scheme-II, respectively) via Eq. \eqref{RL}. 
A schematic representation of the algorithm is shown in 
Fig. \ref{fig:iteration}. Note that at each step, the
scaling goes as $n_Ln_v^2$. With $n_L << n_o^2n_v^2$, our scheme offers significant computational savings over the
conventional CC algorithm.

\section{Results and Discussion :}
In this section, we will discuss the efficiency and 
accuracy of the ADCC formalism and will compare it 
against the canonical coupled cluster method. As a proof-of-concept, without any loss of generality, we will
restrict the cluster amplitudes to only doubles (CCD).
All the results are obtained with our in-house codes and
the convergence threshold was set to be $10^{-6}$. 
Furthermore, no DIIS optimization is used to accelerate
the convergence, even though inclusion of it is fairly straightforward. 

\begin{figure}[!h]
    \centering
\includegraphics[width=\linewidth]{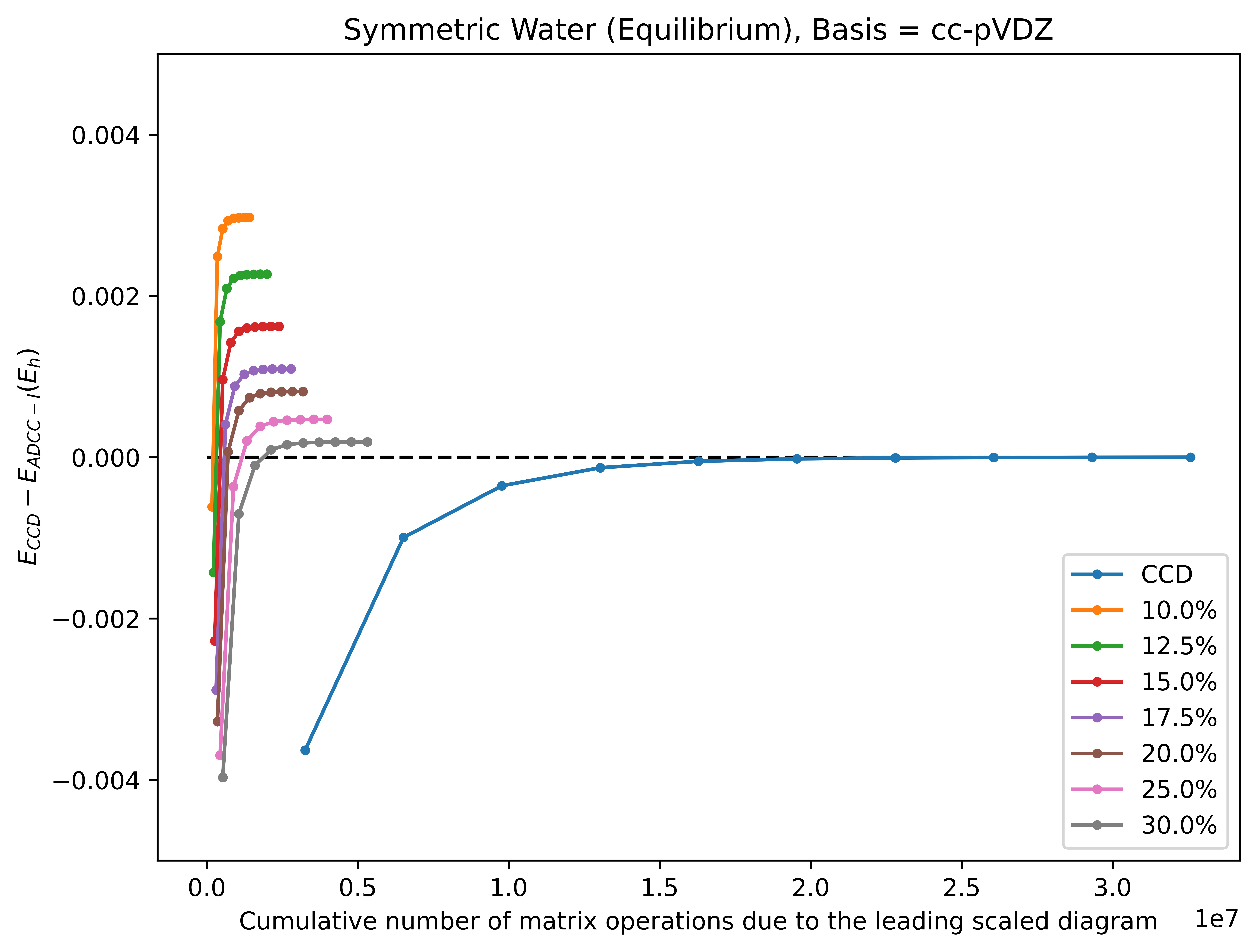}
\caption{Accuracy of ADCCD scheme-I for various size of LAS as function of the cumulative cost of the ADCCD calculation obtained from the highest scaling term. The blue line shows the cumulative number of matrix operation for the construction of the diagram containing all particle contraction for canonical CCD.}
    \label{fig:cumulative}
\end{figure}

As our objective is to simulate the canonical CCD optimization in a 
reduced subspace with order of magnitude less 
computational scaling, it is imperative to take the 
same as our reference. In Fig. \ref{fig:adcc_ccd_plots}, we have plotted the 
difference in energy obtained with ADCCD (both scheme-I 
and scheme-II) and canonical CCD as a function of the 
size of \textit{LAS}. Note that the \textit{percentage of \textit{LAS} (N)}
in the horizontal axis signifies that the largest $N\%$
of total nonzero amplitudes (at the MP2 level) are taken
as independent parameters. Thus, a larger \textit{LAS} signifies
more number of principal amplitudes taken as the 
independent variable, against which all the remaining 
amplitudes (the auxiliary ones) are mapped. Note that 
the horizontal blue line \textit{(y=0)} denotes the 
exact CCD results and the light-blue shaded region
indicates a $0.5$ milli-Hartree ($mE_h$) energy band vertically each sides 
of the reference energy line that provides a measure 
of the accuracy. Four different molecules were chosen 
for our application which 
have distinct variety of electronic complexity: (a) 
symmetric water (equilibrium geometry), (b) asymmetric 
water, (c) ammonia at equilibrium geometry and (d) 
formaldehyde\cite{johnson2006nist}. As we expand \textit{LAS}, the energy
calculated from ADCCD (both scheme-I and scheme-II)
converge monotonically to the corresponding CCD values.
Understandably, ADCCD scheme-I is somewhat better than
ADCCD scheme-II due to the fact that the former takes 
into account the complete feedback coupling of the 
auxiliary amplitudes towards the principal amplitudes, 
while the later does so in an approximate manner. 
For most of the cases, ADCCD Scheme-I enters the 
shaded region in between $20\%$-$25\%$ of \textit{LAS}
and acquires a remarkable 
accuracy of $\sim 0.1 mE_h$ at around $30\%$ in almost 
all the cases under consideration, while ADCCD scheme-II
does so around $25\%$-$30\%$ and $40\%$ of the \textit{LAS} 
percentage, respectively. This implies that with $70\%$
reduction in the number of independent amplitudes, ADCCD scheme-I can achieve an accuracy 
within tens of $\mu E_h$ as compared to the canonical CCD, while 
ADCCD scheme-II does so with $60\%$ reduction in the
number of amplitudes. We will soon analyse how these 
tremendous reduction in the number of independent 
parameters transcends into reduction in the number of
matrix operations needed to achieve the accuracy.

In the inset of each subfigure of Fig. 
\ref{fig:adcc_ccd_plots}, we have plotted the overlap
between the adiabatically determined cluster amplitudes 
and the canonical cluster amplitudes. Towards this, we
have mapped the amplitudes as normalized column vectors
$\ket{t_{CCD}}$ and $\ket{t_{ADCCD}}$. Note that for ADCC 
schemes, both the principal and auxiliary amplitudes are 
included in $\ket{t_{ADCCD}}$ where their 
index-ordering were kept consistent. The accuracy of 
the ADCCD cluster amplitudes are measured by the 
quantity $(1- \bra{t_{ADCCD}}\ket{t_{CCD}})$.  
The inset plot shows the monotonic approach of 
the overlap value to one for scheme-I with increase in the 
size of
\textit{LAS}, signifying that the ADCCD cluster 
amplitudes are of the similar accuracy to those 
obtained by the canonical CCD scheme.

Having discussed the precision of the optimization algorithm, we now 
comment on the number of matrix operations needed. While canonical CCD requires 
$\mathcal{O}(n_o^2n_v^4)$ matrix operations each iterative cycle, our scheme requires
$\mathcal{O}(n_Ln_v^2)$ matrix operations. In Fig.\ref{fig:cumulative}, we have 
demonstrated that one can achieve precision of the order of tens of $\mu E_h$
with LAS taken to be $30\%$ while it takes at least an order fewer number of
cumulative matrix operations. One may note that the ADCC scheme, at each iterative cycle, 
requires one order of magnitude less memory compared to canonical CCD, and thus 
enabling large scale computations without significantly sacrificing the accuracy.

\section{Conclusion and Future Outlook :}
In this article we have rigorously established the 
fact that the discrete-time CC
iteration dynamics can be considered as a collective 
motion of coexisting slow 
and fast relaxing modes, manifesting a dynamical
hierarchical structure. Although the
concepts of adiabatic elimination and slaving 
principle in synergetics and 
nonlinear dynamics are applicable for 
non-equilibrium systems in most of the existing 
literatures, borrowing these ideas we have successfully
developed an equilibrium dynamical formalism 
(in the amplitude space) to study 
the complex transient behaviour of coupled
cluster iteration before convergence. Our
scheme shows order of magnitude reduction in 
computational efforts to obtain molecular ground state 
energy with sub-$mE_h$ to tens of $\mu E_h$ 
precision to the conventional CC schemes. The results 
are, furthermore, systematically improvable with 
inclusion of more number of amplitudes in the \textit{LAS}. 
However while doing so we have neglected 
some higher order post-adiabatic 
terms that opens up scopes for
further improvement of the adiabatic 
results\cite{haken1975higher}.    
Also, so far we have hived off \textit{LAS} from the full 
cluster amplitude space purely based on the observation
that the cluster 
amplitudes have a difference in relaxation timescales with respect to their 
magnitude distribution. There exist some information theoretical techniques 
\cite{haken2016information, haken1985application, haken1987self, haken2006information, haken2021information, haken1985information} and entropy maximization approaches
\cite{haken1986maximum, haken1989synergetics} to accurately determine the 
principal amplitudes. Although most of these principles are mathematically 
heavy and applicable for thermodynamical systems far from 
equilibrium, we are looking for devising our 
own methods based on these concepts to construct \textit{LAS} in a more mathematically 
accurate manner. Some stochastic adiabatic elimination 
principles\cite{schoner1986slaving, schoner1987systematic,aoki2013slow, wu1990relations,constable2013stochastic,li1992relations, wu2000suppressing} can 
also be applied to introduce stochasticity or external noise in CC to study
thermal properties. We can utilise some studies that have shown how the stable 
and unstable manifolds can suddenly change or collide among themselves in the 
phase portrait under certain conditions to show some interesting phenomena\cite{grebogi1983crises,takatsuka2018adiabatic}. 
In the immediate future, we will try to incorporate the 
higher order post-adiabatic correction terms to
investigate further the intricate mathematical structures of the theory 
and to improve the results on top of ADCC. 

\section{Acknowledgments}
RM acknowledges the 
financial support from Industrial Research and
Consultancy Centre, IIT Bombay, and 
Science and Engineering Research Board, Government
of India.

\section*{Data Availability}
The data is available upon reasonable request to the corresponding author.

\section*{Keywords:}  Electronic structure.  Ab initio calculations. Quantum Chemistry. Nonlinear Dynamics. Synergetics.

%aipnum4-2.bst 2019-01-14 (MD) hand-edited version of apsrev4-1.bst

%\section{APPENDIX:}
%
%\begin{multline}
%    H_{S_i}^d = (1 + P(i,j)P(a,b)) \Big(-f_{ii}+f_{aa}+\frac{1}{2}v_{ab}^{ab}+2v_{ia}^{ai}\\-(1+\delta_{ij}\delta_{ab}) v_{ia}^{ia}+ \frac{1}{2}v_{ij}^{ij}- v_{ib}^{ib} \Big)\\
%    S_i = \{ijab\} \in \textcolor{red}{SAS}
%\end{multline}    

%\bibliography{./literature}

\end{document}